\documentclass[aps,pra,twocolumn,showpacs,superscriptaddress,reprint]{revtex4-1}
\usepackage{amsmath}
\usepackage{epsfig}  
\usepackage{color}
\usepackage{endnotes}
\usepackage{natbib}
\usepackage[colorlinks,breaklinks]{hyperref}
\usepackage{nicefrac}
\usepackage{bm}
\usepackage{dcolumn}
\usepackage{graphics}
\usepackage{amsmath}
\usepackage{amssymb}
\usepackage{color}
\newcommand{\bra}{\langle}
\newcommand{\ket}{\rangle}

\newcommand{\bs}[1]{\ensuremath{\boldsymbol{#1}}}

\newcommand{\be}{\begin{equation}}
\newcommand{\ee}{\end{equation}}
\newcommand{\bea}{\begin{align}}
\newcommand{\eea}{\end{align}}
\newcommand{\beqa}{\begin{eqnarray}}
\newcommand{\eeqa}{\end{eqnarray}}
\newcommand{\bk}{{\bs k}}

\begin{document}

\title{Photo Reactions with Universal Trimers}

\date{\today}

\author{Betzalel Bazak}
\email{betzalel.bazak@mail.huji.ac.il}
\affiliation{Racah Institute of Physics, The Hebrew University, 
             91904, Jerusalem, Israel}

\author{Nir Barnea}
\email{nir@phys.huji.ac.il}
\affiliation{Racah Institute of Physics, The Hebrew University, 
             91904, Jerusalem, Israel}

\begin{abstract}
Considering one-body and two-body currents, we study the photoassociation and 
photodissociation of universal bosonic trimers.
Analyzing the relative importance of these currents we identify two physical 
scenarios
(i) Normal hierarchy, where naive power counting holds and the one-body
current dominates, and (ii) Strong hierarchy, where the one-body current is 
suppressed.  
For both scenarios we observe that at the high frequency tail, the response 
function exhibits log periodic oscillations in transition to or from any 
continuum state regardless of the reaction partial wave channel. 
In contrast, near threshold log periodic oscillations appear only 
in the leading $s$-wave components.
These oscillations are the fingerprints of universal Efimov physics.
We discuss the relevance of this effect to contemporary experiments in 
ultracold atoms.
\end{abstract}


\maketitle

{\it Introduction} ---
The low energy physics of two neutral particles interacting via short range 
force depends on a single parameter, the $s$-wave scattering length $a$.
As long as the energy of the system is much smaller than $\hbar^2/M r_0^2$, 
the energy scale associated with the typical range of the potential $r_0$ and
the mass $M$, the properties
of the system are indifferent to the details of the inter particle force.
In this limit the system is said to exhibit universal behavior that is
independent of its actual constituents.

The window for probing universality opens up when the scattering length $a$ is
much larger than the typical range of the potential $r_0$. Such is the situation 
in nuclear physics \cite{BraHam06}, in Helium molecules \cite{GriSchToe00} 
and in magnetically manipulated ultracold atoms \cite{InoAndSte98}.

Under the condition $|a| \gg r_0$ universality can be manifested not only in the
two-body case but also in other few and many-body systems.
The system of three identical bosons is a particular case.     
In the limit of resonating two-body interaction $|a|\longrightarrow\infty$
universality is associated with a series of weakly bound three-body 
states, known as Efimov trimers \cite{Efi70,KraMarWal06}, that appear even when the
interaction supports no bound dimer.
The spectra of these trimers reveals a discrete  
scaling symmetry, 
resulting from quantum mechanical breaking of classical
scaling symmetry, that also yields log periodic oscillations 
in the trimer's wave function.

These log periodic oscillations are the fingerprints of the Efimov effect.
They were predicted to appear in various aspects of universal trimer physics.
One example is the oscillations of the 3-body recombination rate constant
as a function of the scattering length $a$ \cite{EsrGreBur99,BedBraHam00, 
ZacDeiDer09,PolDriHul09,GroShoKok09,GroShoMac11}.
Another examples revealing log periodic oscillations are 
the atom-dimer scattering length \cite{Efi79} and disintegration cross-section,
and the energy dependent collisional recombination rate of Efimov trimers in 
cold gas \cite{WanEsr11}.

In this manuscript we discuss another aspect of the Efimov
effect that have drawn very little attention so far, that is the appearance of
log periodic oscillations in 
{the photo reaction cross section}
of universal trimers.
To this end we consider one-body and two-body currents and analyze their
relative importance.

{\it Photo Reactions} ---
The photodissociation and the photoassociation cross sections are both related
through kinematic factors to the photo response function 
 \be\label{resp1}
  S(\omega) = \bar{\sum_i}\sum_f |\bra  \Psi_f | \hat{H}_I | \Psi_B, \bs{k} \lambda 
 \ket|^2 \delta(E_f-E_B-\hbar\omega)  
 \ee
that describes the transition of the trimer from a bound state $\Psi_B$ with
binding energy $E_B$ into a continuum state $\Psi_f$ with energy $E_f$ by absorbing 
a photon of momentum $\bs{k}$, polarization $\lambda$, and energy $\hbar\omega=\hbar k c$.
$\sum_f$ stands for integration over the final states
and $\bar\sum_i$ averages over the appropriate initial states.

The coupling between neutral particles and radiation field takes 
the form
$\hat{H}_I=-e\int d \bs x \bs \mu({\bs x})\cdot \nabla \times \bs A(\bs x)$
where $\bs \mu$ is the magnetization density and $\bs A$ is 
the electromagnetic field.
In effective low energy theory
\cite{Kap05}  
the magnetization density contains not only one-body current, but also
two-body and more body currents
\be
\bs \mu(\bs x)=\bs \mu^{(1)}(\bs x)+\bs \mu^{(2)}(\bs x)+ \ldots
\ee
Using low momentum expansion, the operators in the theory can be arranged
in powers of $k/\Lambda$, and $Q/\Lambda$, 
where $Q$ is the
typical particle momentum of the system under consideration.
$\Lambda$ is the cutoff momentum of the theory, it
reflects the point $\sim\hbar/r_0$ at which short range physics, ignored by
the low energy theory, is becoming important.
At the leading order (LO) and next to leading order (NLO) 
the one-body current takes the form \cite{SonLazPar09} 
\be
\bs\mu^{(1)}(\bs x)= \sum_j\bs s_j\left(\mu_0+L_1
\frac{\bs{k}^2}{\Lambda^2}\right) \delta(\bs x-\bs r_j),  
\ee
where $\mu_0$ is the magnetic moment of a single particle, 
$\bs{r}_j$, $\bs{s}_j$ are the position and spin of particle $j$,
and $L_1$ is a shape parameter describing the particle's form factor. 
The two-body contribution to the current enters at the next order (N$^2$LO),
or $(Q/\Lambda)^3$, and takes the form \cite{SonLazPar09}
\be\bs\mu^{(2)}(\bs x)= \sum_{i<j}(\bs s_i+\bs s_j) 
\frac{L_2}{\Lambda^3} \delta(\bs x-\frac{\bs r_i+\bs r_j}{2})
       \delta_{\Lambda}(\bs r_i-\bs r_j)\;. 
\ee
The low energy parameter $L_2$ is the coupling constant between the radiation
field and the four boson fields. Its value can be fixed studying dimer
photoreactions. The notation $\delta_{\Lambda}(\bs r)$ stands for Dirac's
$\delta$-function smeared over distance $\hbar/\Lambda$.

Three-body currents, associated with two more bosonic fields are
suppressed by another factor of $(Q/\Lambda)^3$ relative to the N$^2$LO
two-body current. Consequently, in the low energy limit $Q\ll\Lambda$ they
can be ignored.

For the bosonic system we assume that the initial and final state 
wave functions can be written as 
a product of a symmetric spin $|\chi\ket$ and 
configuration space $|\psi\ket$ components, $|\Psi\ket=|\psi\,\chi\ket$. 
We shall further assume that the spin component of the
wave function is frozen throughout the photoreaction process, thus
$|\chi_f\ket=|\chi_B\ket$.
Using these assumptions, 
the one-body transition matrix element in (\ref{resp1}) takes the form
\begin{align}\label{hi1b}
\bra \Psi_f | & \hat H_I^{(1b)} | \Psi_B,\bk \lambda \ket=
\\ &
-i\sqrt{\frac{\hbar c^2}{2V\omega_k}} 
 \left(\mu_0+L_1\frac{\bk^2}{\Lambda^2}\right)\bra \bs s \ket  \sum_{j=1}^3 
      \bra \psi_f | e^{i\bs k \cdot \bs r_j}| \psi_{B} \ket \;,
\nonumber
\end{align}
whereas the two-body reads
\begin{align}\label{hi2b}
\bra &\Psi_f |\hat H_I^{(2b)} | \Psi_B,\bk \lambda \ket=
\\ &
-i\sqrt{\frac{\hbar c^2}{2V\omega_k}} 
   \frac{L_2}{\Lambda^3}{2}\bra \bs s \ket  \sum_{i<j}^3 
      \bra \psi_f | e^{i\bs k \cdot (\bs r_i+ \bs r_j)/2}
             \delta_{\Lambda}(\bs r_i- \bs r_j)| \psi_{B} \ket 
      \;.
\nonumber 
\end{align}
Here $\bra \bs s \ket = \frac{1}{3}\sum_j \bra \chi_{B} |
\bs s_j \cdot (\bk \times \hat e_{\bk \lambda}) | \chi_{B} \ket $, 
and we have used box normalization of volume $V$.

The appearance of log periodic oscillations in the high frequency tail of
spin flip rf reactions in ultracold atoms was recently predicted by Braaten
{\it et} al. \cite{BraKanPla11}. Studying rf reactions in the unitary limit
$a\longrightarrow\infty$ they used effective 
field theory methods to analyze the Franck-Condon factor dominating 
photo induced spin-flip reactions and found that in the limit
$\omega\longrightarrow\infty$ the 3-body response function takes the form
$   (A_1+A_2\sin[s_0\ln(\hbar\omega/|E_B|)+2\phi])/(2m\omega^2),$
where $A_1,A_2,\phi$ are constants, and $s_0\approx 1.00624$ is the Efimov
parameter. 

For frozen spin reactions the Franck-Condon term is zero,
nonetheless, under the assumptions leading
to Eqs. (\ref{hi1b})-(\ref{hi2b}), 
we will show in the following that
(i) In the unitary limit the high frequency tail of the response 
function contains log periodic oscillations in all partial waves. 
(ii) The leading $s$-wave component of the response function exhibits 
log periodic oscillations throughout the spectrum, from threshold to infinity. 


In the long wavelength limit, the exponent in the LO one-body matrix
element, Eq. (\ref{hi1b}), can be expanded to yield
\begin{align} \label{expand}
\sum_{j=1}^3 e^{i\bs k \cdot \bs r_j} &\approx  3
+i\sum_{j=1}^3 \bk \cdot \bs r_j
-\frac{1}{6} \sum_{j=1}^3 k^2 r_j^2 \cr &
-\frac{4\pi}{15} \sum_{j=1}^3 k^2 r_j^2 \sum_m Y_{2 -m}(\hat{k}) Y_{2
  m}(\hat{r}_j) + \ldots, 
\end{align}
where $Y_{lm}$ are the spherical harmonics.
The zeroth order operator is just the Franck-Condon factor which 
dominates the spin-flip reaction \cite{TscRit11}, but has no contribution in
frozen-spin reactions.
The first order operator is the dipole, which for identical particles is
proportional to the center of mass and hence does not affect the relative 
motion of the particles. 
Consequently, the one-body current is dominated by the
$O\left((kr)^2\right)$ quadrupole and $r^2$ operators 
\cite{BazLivBar12,LivBazBar12}.

Assuming that the photon energy is of the order of the binding energy $E_B$
and observing that $r\approx \sqrt{\hbar^2/M E_B}$, the long wavelength
expansion parameter can be written as $k r \approx \sqrt{E_B/Mc^2}\approx Q/Mc $.
Comparing $Q/Mc$ to the low energy expansion parameter $Q/\Lambda$, we see that
the importance of the two-body current depends on the relative magnitude 
of the two high momentum scales $\Lambda$ and $Mc$. If the scales are
such that $Q/\Lambda \ll {\Lambda}/Mc$ the two-body currents appearing at order
$(Q/\Lambda)^3$ is expected to be much smaller than the second order 
$(kr)^2\approx (Q/Mc)^2$ terms. Under this circumstances, 
we have \emph{normal hierarchy} as suggested by naive power
counting. This is the situation in any effective low energy theory in the
limit $Q \longrightarrow 0$. The electro-magnetic currents in nuclear physics
are a fine example of this hierarchy, see e.g. \cite{SonLazPar09,PasGirSch11,RozGolKol11}.
On the other hand, when the short range energy scale is much smaller 
than the mass scale  $\Lambda \lll Mc$, we can face a situation where the 
one-body current proportional to $(Q/Mc)^2$ becomes negligible in comparison 
to the two-body current.
This case of \emph{strong hierarchy} is typical for frozen-spin experiments 
in ultracold atoms \cite{LomOttSer10,NakHorMuk11,MacShoGro12}.
There, the ratio between the scattering length and the effective range is much 
smaller than the separation between the mass scale and the binding energy.

In the following, after addressing the three-body wave function in the unitary
limit, we would analyze these two limiting physical scenarios of normal  
and strong hierarchies. 

{\it The three-body system} ---
The dynamics of a quantum 3 particle system is governed by the Hamiltonian 
$H=T+U$, where $T$ is the kinetic energy operator and the 
potential $U$ is a sum of 2 and 3-body forces.
Removing the center of mass coordinate, the system can be described by
the Jacobi vectors
$\bs x=\sqrt \frac {1}{2} (\bs r_2-\bs r_1)$, and 
$\bs y = \sqrt \frac {2}{3}\left(\bs r_3 - \frac{\bs r_1+\bs r_2}{2}\right)$,
which we transform into 
the hyperspherical coordinates $(\rho,\Omega)$.
$\rho^2=x^2+y^2$, $\Omega=(\alpha,\hat x, \hat y)$ and
$\tan \alpha=x/y$.

For low energy physics, when the extension of the wave function is
much larger than the range of the 2-body potential, one can utilize the zero range
approximation, where the action of the potential is represented through
appropriate boundary conditions when two particles approach each other
\cite{SieJen87}.
In a similar fashion, the short range 3-body force can be replaced by the 
boundary condition $\partial \log \psi/\partial \log\rho=C$ at $\rho=\rho_0$. 
Here we shall assume a hard core potential, i.e. 
$C \longrightarrow\infty$.
The cutoff hyperradius $\rho_0$ is our three body parameter, 
and it can be fitted for example to the trimer's binding energy.


A remarkable aspect of the zero range approximation in the unitary limit
is the factorization of the wave function into a product of hyperangular
and hyperradial terms \cite{Mac68},
$\psi(\rho,\Omega)=\rho^{-5/2}\mathcal R(\rho)\Phi_{\nu}(\Omega)$ .
The hyperangular channel functions $\Phi_{\nu}(\Omega)$ are the solutions of 
the adiabatic hyperangular equation with eigenvalue $\nu^2$.
The hyperradial functions $\mathcal R(\rho)$ are the solutions 
of the hyperradial equation,
\begin{align} \left (-\frac{\partial^2}{\partial\rho^2}+\
        \frac{\nu^2-1/4}{\rho^2}\right)\mathcal R(\rho)
         = \epsilon\mathcal R(\rho)\;,
\end{align}
subject to the boundary condition $\mathcal R(\rho_0)=0$. Here 
$\epsilon=2ME/\hbar^2$. 
The lowest eigenvalue of the adiabatic hyperangular equation appears in
the $s$-wave channel $\nu_0=is_0$, all other eigenvalues are positive. 
The Efimov effect \cite{Efi70} results from the fact that $\nu_0$ is imaginary. 
It is limited to the $s$-wave as there are no imaginary eigenvalues of 
the hyperangular equation in any other channel.


Substituting ${\cal R}=\sqrt{\rho}u(\rho)$, the hyperradial equation is just
the Bessel equation. The bound state solution for the lowest channel
is proportional to the modified Bessel function of the second kind 
and imaginary order, $\sqrt \rho K_{i s_0}(\kappa \rho)$, 
where $\kappa=\sqrt {-\epsilon}$. 
The value of $\kappa$ is fixed by the 3-body boundary condition at $\rho_0$.
The result is the discrete Efimov spectrum,
$\epsilon_n/\epsilon_0=e^{-2\pi n/s_0}\approx 515^{-n} $.
The normalized wave functions are 
\be \label{bound_wf}
\mathcal R^{(n)}_B(\rho)=N_B\kappa_n \sqrt \rho K_{i s_0}(\kappa_n \rho),
\ee
where $N_B=\sqrt{2 \sinh (s_0 \pi)/s_0 \pi}.$

For scattering states $\epsilon>0$ the hyperradial wave function
is composed of the Bessel functions of the
first and second kind of order $\nu$, and the real part of these functions
if $\nu$ is imaginary,
\be \label{cont_wf}
\mathcal R_f(\rho)=\sqrt{\frac{q \rho N_s}{R}} 
\Big( \sin \delta \,\mathrm{Re}[J_\nu(q \rho)] +
\cos \delta \,\mathrm{Re}[Y_{\nu}(q \rho)] \Big),
\ee
where $q=\sqrt {\epsilon}$. Here we assume normalization in a sphere of 
radius $R$, and $N_s=1/2$ ($\pi$) for imaginary (real) $\nu$.
The phase shift $\delta$ is to be found from the boundary condition, 
$\mathcal R_f(\rho_0)=0$.


{\it Normal hierarchy} ---
For the normal hierarchy case, the one-body current
dominates the photo reaction and the transition matrix element
is given by (\ref{hi1b}). It
can be written as a power series in the 
hyperradius $\rho$ in the following form 
$\beta_1\sum_j e^{i\bs k\cdot \bs r_j}=\sum_{m=0,\mu}^\infty 
      {\cal A}_{m \mu} \rho^{m}{\cal Y}_{\mu}(\Omega) $,
where $\beta_1$ is a prefactor that can be deduced from (\ref{hi1b}), ${\cal
  Y}_{\mu}(\Omega)$ are the hyperspherical harmonics \cite{Ave89}
that span the hypersphere $\Omega$, and
${\cal A}_{m \mu}$ are the expansion coefficients.
Substituting this expansion into (\ref{hi1b}), the transition matrix element
between a bound $s$-wave state and a continuum state is
\be \label{power}
 \bra \psi_f | \beta_1\sum_j e^{i\bs k\cdot \bs r_j} | \psi_B \ket =
   \sum_{m \mu} {\cal A}_{m\mu} C_{\nu\nu_0}^\mu
    \int d\rho {\cal R}_f^*\rho^m {\cal R}_B \;,
\ee

where $C_{\nu \nu_0}^\mu =\bra \Phi_{\nu}|{\cal Y}_{\mu}|\Phi_{\nu_0}\ket$ are
the hyperangular matrix elements.
Here and in the following we omit the trimer's excitation index $n$ 
since the discussion applies to any of these states. 

Substituting Eqs. (\ref{bound_wf},\ref{cont_wf}), the 
matrix elements are proportional to integrals of the type
\be \label{IJm}
 {\cal I}_{J}(\nu,m) = \int_{\rho_0}^\infty d\rho\, \mathrm{Re}[J_{\nu}(q \rho)] 
\rho^{m+1} K_{i s_0}(\kappa \rho).
\ee
or ${\cal I}_{Y}(\nu,m)$ with $Y_{\nu}$ replacing $J_{\nu}$. 
We note that the bound
state is invariant to the Efimov scaling $\rho_0\longrightarrow e^{-\pi/s_0}\rho_0$, 
therefore we can replace the lower limit of the integral by zero \cite{Note}.
Evaluating the integral we get
\be \label{IJmx}
 {\cal I}_J(\nu,m) = {\rm Re}\left[
\frac{2^m N_{\nu_0,\nu}^m}{\kappa^{m+2}}\left(\frac {q}{\kappa}\right)^\nu
{}_2F_1\left(a,b;c;-(q/\kappa)^2\right)\right]
\ee
where $_2F_1$ is the hypergeometric function with parameters
$a= \frac{m-\nu_0+\nu}{2}+1$, $ b=\frac{m+\nu_0+\nu}{2}+1$, and $c=\nu+1$,
and $N_{\nu_0,\nu}^m=\Gamma(a)\Gamma(b)/\Gamma(c)$. 
Similar results are obtained for ${\cal I}_{Y}$.

In the limit of reaction close to threshold, $q\ll\kappa$, this expression can
be approximated by, 
\be
{\cal I}_J(\nu,m) \approx {\rm Re}\left[
\frac{2^m N_{\nu_0,\nu}^m}{\kappa^{m+2}}\left(\frac {q}{\kappa}\right)^\nu
\right].
\ee

For $s$-wave transition the particles in the final state may move
along the $\nu=\nu_0$ adiabatic potential and since
$
(q/\kappa)^{\nu_0}=
\cos(s_0\ln \frac{q}{\kappa})+i\sin(s_0\ln \frac{q}{\kappa}),
$
the response function acquires log periodic oscillations.
The remarkable aspect of this result is the fact that in the long wavelength 
limit the $s$-wave transition operator $r^2$ is the dominant multipole, and
therefore they can be observed, in principle, studying photoreactions of Efimov
trimers close to threshold. 

In the limit of large energy transfer where $q\gg\kappa$, but still smaller than 
the cutoff momentum $\Lambda$, the reaction probes the short range part of the 
trimer's wave function and the transition integral takes the form
\be \label{largeq}
{\cal I}_J(\nu,m) \approx
\frac{2^{m+1}}{q^{m+2}}
 \mathrm {Re}\left[\frac{\Gamma (\nu_0) 
   \Gamma \left(\frac{m-\nu_0+\nu}{2}+1\right)}{\Gamma
   \left(\frac{-m+\nu_0+\nu}{2}\right)}\left(\frac{q}{\kappa}\right)^{\nu_0}\right]\;.
\ee
This result indicate that for any $\nu$ the transition matrix element exhibits
log periodic oscillations in the large energy tail, attenuated by $q^{-m-2}$. 
These oscillations reflect the structure of the Efimov trimers at short
distances, Eq. (\ref{bound_wf}). Consequently, the number of oscillations will
reflect the number of nodes in the trimer's wave function. 

At this point we would like to study in some detail the response function
associated with the $r^2$ operator. As we have already indicated, for normal 
hierarchy this is the dominant term at low photon energies \cite{BazBar14}.
The $r^2$ operator is extremely simple,
 $\sum_j r_j^2 = \rho^2 + 3R_{CM}^2$ where $R_{CM}$ is the center 
of mass radius, and therefore 
\be \label{r2me}
  \bra \psi_f | - \frac{\beta_1}{6}\sum_j k^2r_j^2 | \psi_B \ket
  = -\frac{\beta_1 k^2}{6} N_B \kappa\sqrt{\frac{q N_s}{R}} {\cal I}(\nu_0,2)\;,
\ee
where
${\cal I}(\nu_0,m)={\cal I}_{J}(\nu_0,m) \sin \delta +  {\cal I}_{Y}(\nu_0,m)
\cos \delta$ 
is the sum of the $J$ and $Y$ transition matrix elements 
${\cal I}_{J}, {\cal I}_{Y}$ with $m=2$. 

To better understand our results, we evaluate the asymptotic expressions
for the reaction matrix elements at threshold and at the high energy tail.
In the limit of $q \ll \kappa$ we obtain
\begin{align}\label{IJYlow}
{\cal I}_{J}(\nu_0,2) &\approx \frac{B_1}{\kappa^4} \cos(s_0 \ln \frac{q}{\kappa}+\phi)
\end{align}
where $B_1=4|\nu_0+1|\approx 5.6745$,
$\phi=\tan^{-1}s_0\approx 0.78851$.
The phase shift is also a log periodic function, 
consequently, near the threshold the matrix element can be well approximated by
\be
{\cal I}(\nu_0,2) \approx 1+\frac{B_2}{2}\cos(2s_0\ln\frac{q}{\kappa})
\ee
where $B_2 \approx 8.475\%$ is the normalized amplitude of these oscillations.

Similarly, in the limit $q \gg \kappa$, one gets 
\begin{align} \label{IY2}
{\cal I}_{J}(\nu_0,2) & \approx -\frac{B_1}{q^4} 
\sin(s_0 \ln \frac{q}{\kappa} +\phi) 
\end{align}
However, in this limit $\delta \approx \pi/4 - q\rho_0$,
and the log periodic oscillations are masked by these linear oscillations.
This situation will certainly complicate an experimental attempt to probe the
log periodic oscillations in this limit.

\begin{figure}\begin{center}
\includegraphics[width=7.6 cm]{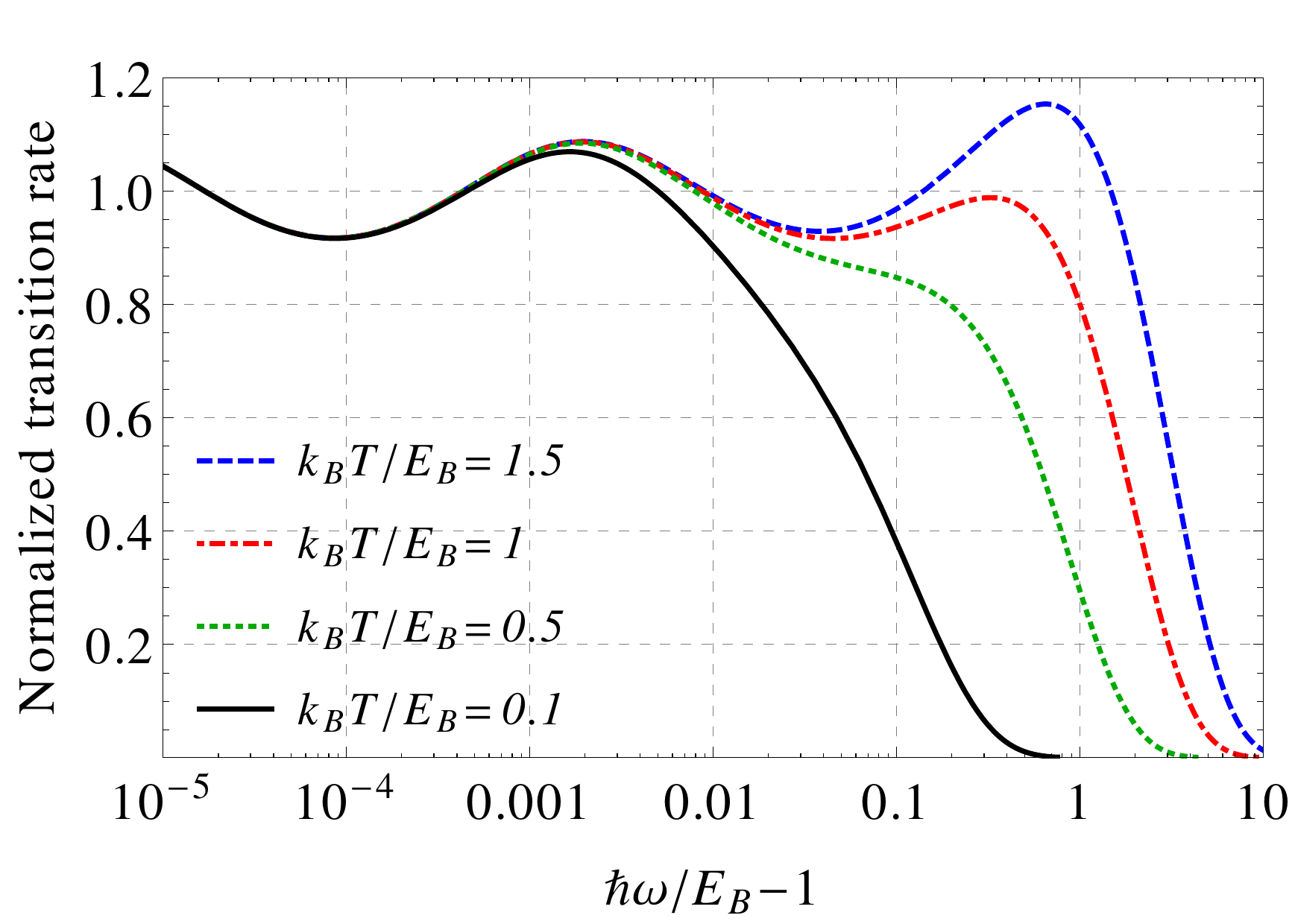}
\caption{\label{3BTlog} (Color online) Normalized trimer photoassociation 
   rate as a function of rf photon frequency for different gas temperatures.
   Blue dashed line $k_BT=1.5 E_B$, red dot-dashed line $k_BT=E_B$, green dotted
   line $k_BT=0.5 E_B$, and black line $k_BT=0.1 E_B$.}
\end{center}
\end{figure}

Substituting the results (\ref{r2me})-(\ref{IY2}) in (\ref{resp1}) we get a
closed form expression for the response function. 
Utilizing this response function, the trimer photoassociation rate 
in a gas of temperature $T$ can be easily calculated by
averaging the response weighted with the probability $P(q)$
of finding three particles in the appropriate continuum state.
Assuming the system is in thermal equilibrium with temperature $T$, 
higher than the condensation temperature, 
$P(q)=\frac{1}{\mathcal{Z}}\frac{R}{\pi}e^{-\beta \hbar^2 q^2 / 2M}$
\cite{BazBar14}. In Fig. \ref{3BTlog} we present 
the resulting transition rates for gas with temperatures 
$k_BT=1.5 E_B, \;k_BT=E_B,\; k_BT=0.5 E_B$, and $ k_BT=0.1 E_B$. 
It can be seen that the number of visible peaks in the trimer photoassociation 
rate depends on the gas temperature, as the 
oscillations are suppressed for $\hbar\omega\geq k_B T$ by the Boltzmann
factor. In the scale  
presented here the log periodic oscillations at the high frequency tail are 
unobservable. 
In reality the lifetime of the trimer may be short and the oscillations at 
energies smaller than the trimer energy width may be smeared.


{\it Strong hierarchy} ---
In case of strong hierarchy where 
$\Lambda \lll M$ and $(Q/\Lambda)^3 \gg (Q/M)^2$
the subleading two-body current becomes dominant and the LO one-body current
is suppressed and becomes negligible. The transition matrix element 
(\ref{resp1}) is then dominated by the leading two-body term (\ref{hi2b}) 
$\beta_{2}\bra\psi_f|\sum_{i<j}\delta_{\Lambda}(\bs r_i-\bs r_j)|\psi_B\ket=
3\beta_{2}/\sqrt{8}\bra\psi_f|\delta_{\Lambda}(\bs x)|\psi_B\ket$, with 
$\beta_{2}=-i\sqrt{\frac{\hbar c^2}{2V\omega_k}} 
   \frac{L_2}{\Lambda^3}{2}\bra \bs s \ket $. 

The behavior of the wave function
when two particles approach each other is governed by the contact interaction. 
In particular, in the limit $x\longrightarrow 0$
$\psi \approx \rho^{-5/3}{\cal R}(\rho)\left(\sin{(\frac{\pi\nu_0}{2})}/2\alpha
+O(1)+O(\alpha)+\ldots\right)$. 
Utilizing this result we get 
\begin{align}\label{transition2b}
\bra \Psi_f |&H_I|\Psi_B,\bs k \lambda\ket = \int_0^{\infty}d\rho\frac{1}{\rho}
{\cal R}_f^*(\rho){\cal R}_B(\rho) \cr &
\times\left(A\Lambda^2|\sin(\pi\nu_0/2)|^2
+O(\Lambda)+O(1)+\ldots\right)\;,
\end{align}
At inter particle distance $r\approx \hbar/\Lambda$ 
the zero range approximation breaks down. Nevertheless,
the details of the wave function at distance $r\leq \hbar/\Lambda$, as well as 
the details of the regulator, i.e. the exact form of $\delta_{\Lambda}(\bs x)$, 
do not affect the structure of the solution. 

The hyperradial integral in (\ref{transition2b}) is nothing but 
${\cal I}(\nu_0,-1)={\cal I}_{J}(\nu_0,-1) \sin \delta +  {\cal I}_{Y}(\nu_0,-1)
\cos \delta$ given by (\ref{IJmx}). Comparing this result 
with Eq. (\ref{r2me}), we conclude that near threshold $q \ll \kappa$ 
the behavior of the photoreaction cross-section is the same for both the strong
and the normal hierarchy cases.

Considering now higher partial waves, we can utilize the completeness of the
hyperspherical harmonics and expand the exponent in
(\ref{hi2b}) into a power series in $\rho$ and hyperspherical harmonics,
following the same arguments leading to Eq. (\ref{power}).
The resulting transition matrix elements are again proportional to
the ${\cal I}_J, {\cal I}_Y$ integrals (\ref{IJm}). Consequently, log periodic 
oscillations will appear in the high energy tail of the cross-section, 
in the same manner as in the normal hierarchy case.

\emph{Summary} ---
Summing up, we have explored photo reactions in universal Efimov
trimers. We analyzed two physical scenarios, (i) Normal hierarchy,
where naive power counting holds and two-body currents are suppressed with
respect to the LO one-body terms, (ii) Strong hierarchy, where the power
counting is distorted and two-body currents dominate.  
For both scenarios we have observed that in all partial waves the 
high energy tail of the
response function exhibits log periodic oscillations. In contrast, 
at threshold log periodic oscillations appear only in the leading $s$-wave 
multipole. These oscillations are the manifestation of the Efimov effect in 
photo reactions.
Considering photoassociation reactions in ultracold atomic gases we have
concluded that contemporary frozen-spin reactions fall into the category of
strong hierarchy. We have found that for temperatures of the order of magnitude 
of the trimer's binding energy the oscillations appear has a pronounced 
structure that can hopefully be observed experimentally without fine tuning. 

 We acknowledge constructive discussions with Chen Ji, Daniel Phillips, 
 Evgeny Epelbaum, and Winfried Leidemann. 
 This work was supported by the ISRAEL SCIENCE FOUNDATION 
 (Grant No.~954/09).

%


\begin{thebibliography}{26}
\bibitem{Note}
  For $q < \kappa$ this approximation amounts to an error 
  $\Delta{\cal I}/{\cal I} \sim x_n^{m+2}$, where $x_0\approx 0.06$ is the 
  largest zero of $K_{i s_0}(x)$ and $x_n$ is the $n$'th root.
\end{thebibliography}
\end{document}